\documentclass[epj-spec]{svjour}
\usepackage{graphics}
\begin{document}
\title{Dijet and photon-jet correlations \\
in proton-proton collisions at RHIC
}
\author{Antoni Szczurek\inst{1}\inst{2}\fnmsep\thanks{\email{antoni.szczurek@ifj.edu.pl}}
 , Tomasz Pietrycki\inst{1}, Anna Rybarska\inst{1} and Gabriela 
\'Slipek\inst{1} }
\institute{Institute of Nuclear Physics PAN, PL-31-342 Cracow, Poland 
\and University of Rzesz\'ow, PL-35-959 Rzesz\'ow, Poland}
\abstract{
We discuss correlations in azimuthal angle as well as 
correlations in two-dimensional space of transverse momenta of two jets
as well as photon and jet.
Some $k_t$-factorization subprocesses are included for the first
time in the literature.
Different unintegrated gluon/parton distributions are used in
the $k_t$-factorization approach. The results depend on UGDF/UPDF used.
The collinear NLO $2 \to 3$ contributions dominate over
$k_t$-factorization cross section at small relative azimuthal
angles as well as for asymmetric transverse momentum configurations.
}
%
\maketitle
\section{Introduction}
\label{intro}

The subject of jet correlations is interesting in the context
of recent detailed studies of hadron-hadron correlations
in nucleus-nucleus \cite{RHIC_correlations_nucleus_nucleus}
and proton-proton \cite{RHIC_correlations_proton_proton} collisions.
Effects of geometrical jet structure were discussed recently
in Ref.\cite{Levai}. No QCD calculation of parton radiation was performed
up to now in this context. Before going into hadron-hadron correlations it
seems indispensable to better understand correlations between jets
induced by the QCD radiation.
Here we discuss the case of elementary hadronic collisions. 
Our analysis is a first step towards the nuclear case. 

In leading-order collinear-factorization approach jets are produced
back-to-back. These leading-order jets are therefore not included into
correlation function, although they contribute a big ($\sim
\frac{1}{2}$) fraction to the inclusive cross section.
The truly internal momentum distribution
of partons in hadrons due to Fermi motion (usually neglected in
the literature) and/or any soft emission would lead to a decorrelation
from the simple kinematical configuration.

In the fixed-order collinear approach only next-to-leading order terms
lead to nonvanishing cross sections at $\phi \ne \pi$ and/or
$p_{1,t} \ne p_{2,t}$ (moduli of transverse momenta of outgoing partons).
In the $k_t$-factorization approach, where transverse momenta
of gluons entering the hard process are included explicitly,
the decorrelations come naturally in a relatively easy to calculate way.
In Fig.\ref{fig:kt_factorization_dijets_diagrams} we show diagrams
included in our calculations which illustrate the physics situation.
The soft emissions, not explicit
in our calculation, are hidden in model unintegrated parton (gluon)
distribution functions (UPDF,UGDF). In our calculation UGDFs or UPDFs
are assumed to be given and are taken from the literature.

The $k_t$-factorization was originally proposed for heavy quark
production \cite{original_kt_factorization}.
In recent years it was used to describe several high-energy processes,
such as total cross section in virtual photon - proton scattering
\cite{UGDF_HERA}, heavy quark inclusive production \cite{BS00,LSZ02},
heavy quark -- heavy antiquark correlations \cite{LS04,LS06},
inclusive photon production \cite{LZ05_photon,PS07_inclusive}, 
inclusive pion production \cite{szczurek03,CS05}, Higgs boson
\cite{Higgs} or gauge boson \cite{KS04} production and dijet
correlations in photoproduction
\cite{SNSS01} and hadroproduction \cite{LO00}.



Up to now no theoretical calculation for photon-jet were presented in
the literature, even for elementary collisions.
In leading-order collinear-factorization approach
the photon and the associated jet are produced back-to-back.
If transverse momenta of partons entering the hard process are
included, the transverse momenta of the photon and the jet are no
longer balanced and finite (non-zero) correlations in a broad range of
relative azimuthal angle and/or in lengths of transverse momenta of 
the photon and the jet are obtained. The finite correlations can be also
obtained in higher-order collinear-factorization approach
\cite{Berends}.

Here we discuss the region of relatively semi-hard jets/photons, i.e.
the region related to the recently measured hadron-hadron
correlations at RHIC and photon-hadron correlations
being analysed \cite{JanRak}. Here the resummation effects
may be expected to be important. The resummation physics is addressed
in our case through the $k_t$-factorization approach.

This presentation is based on recent publications of the authors
\cite{SRS07,PS07}. Here we discuss only the main idea and present some
representative results. More details can be found in Refs. \cite{SRS07,PS07}.

\section{Formalism}


\begin{figure}    
\begin{center}
\resizebox{0.35\columnwidth}{!}{%
\includegraphics{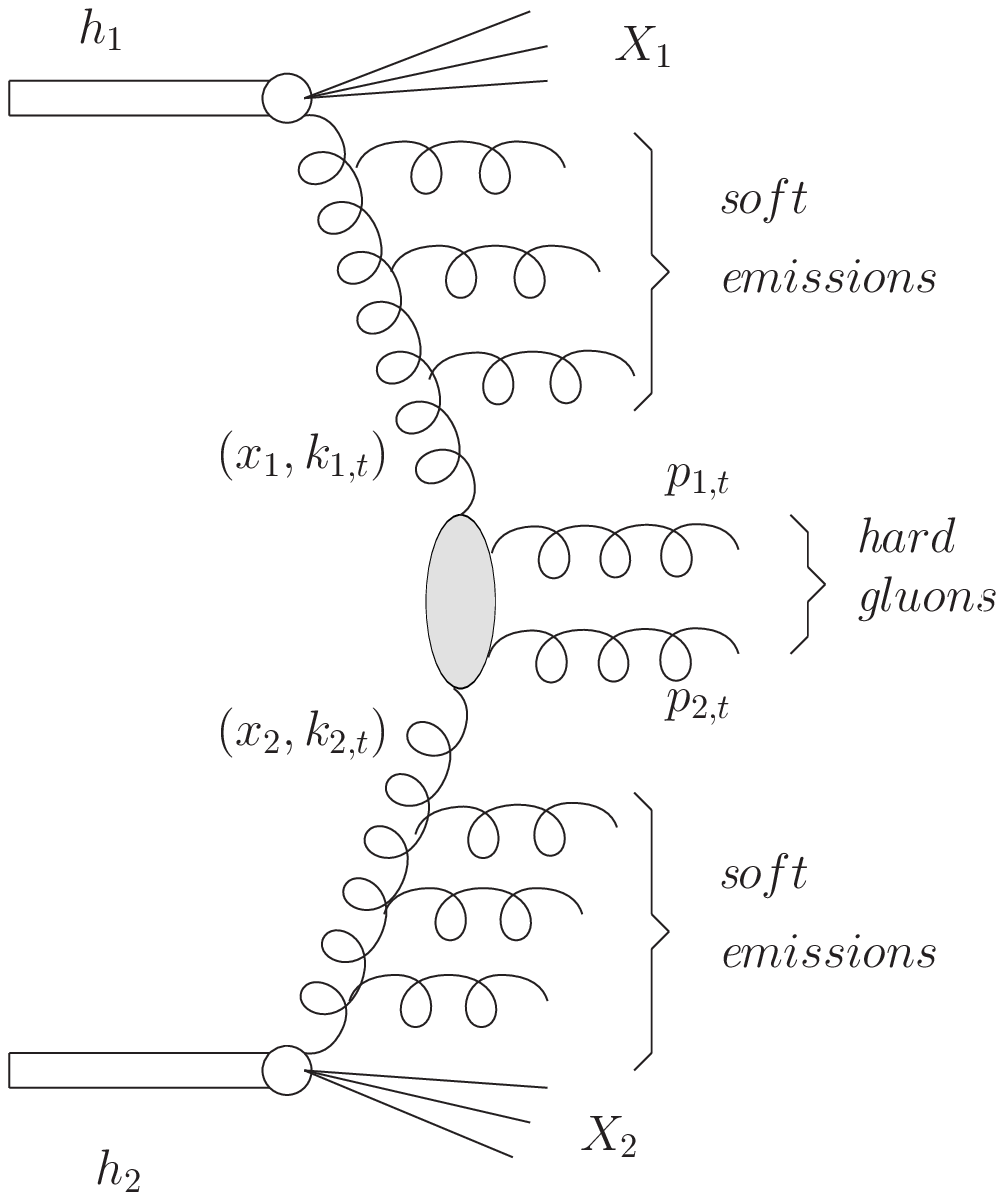} }
\resizebox{0.35\columnwidth}{!}{%
\includegraphics{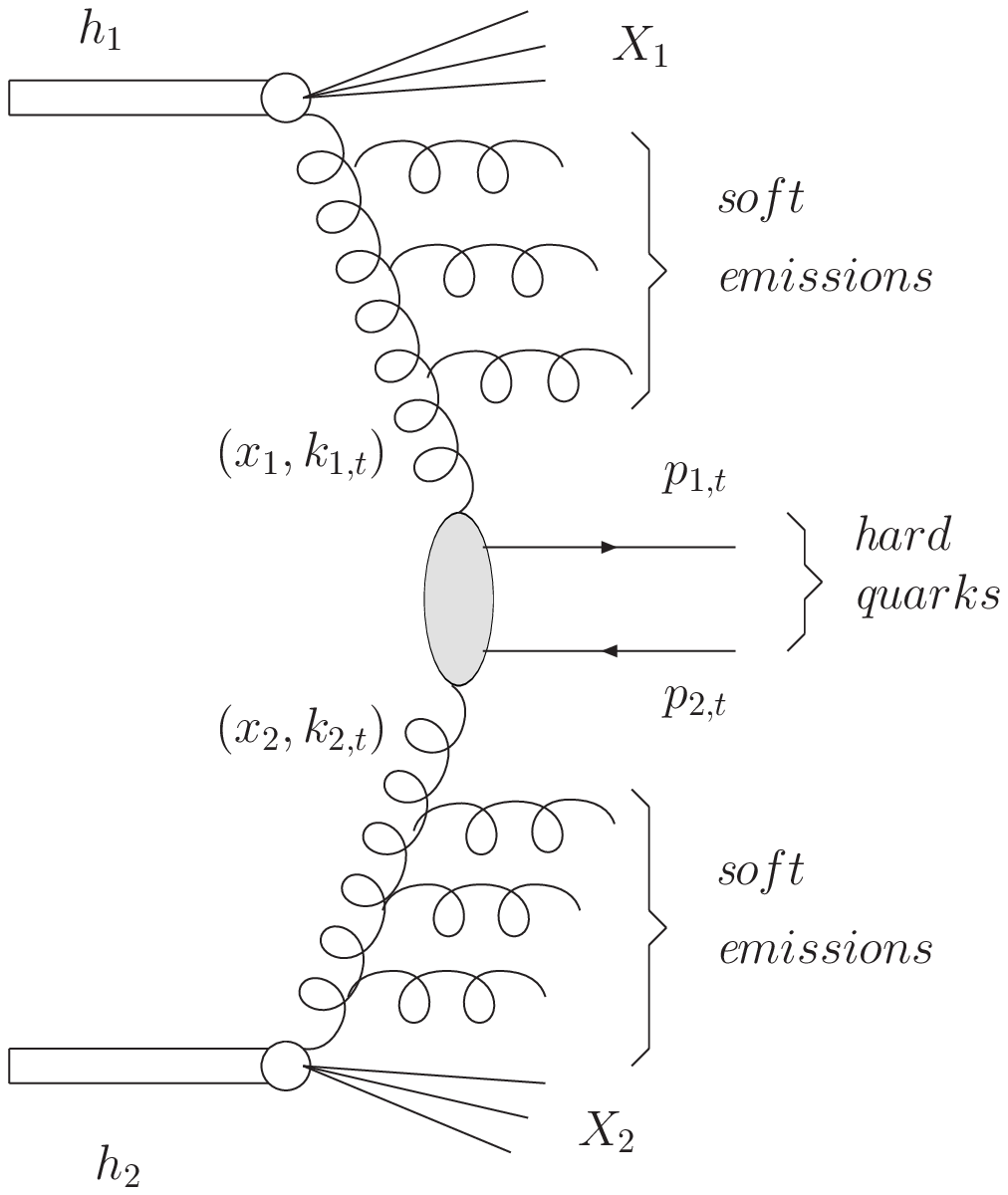} }
\resizebox{0.35\columnwidth}{!}{%
\includegraphics{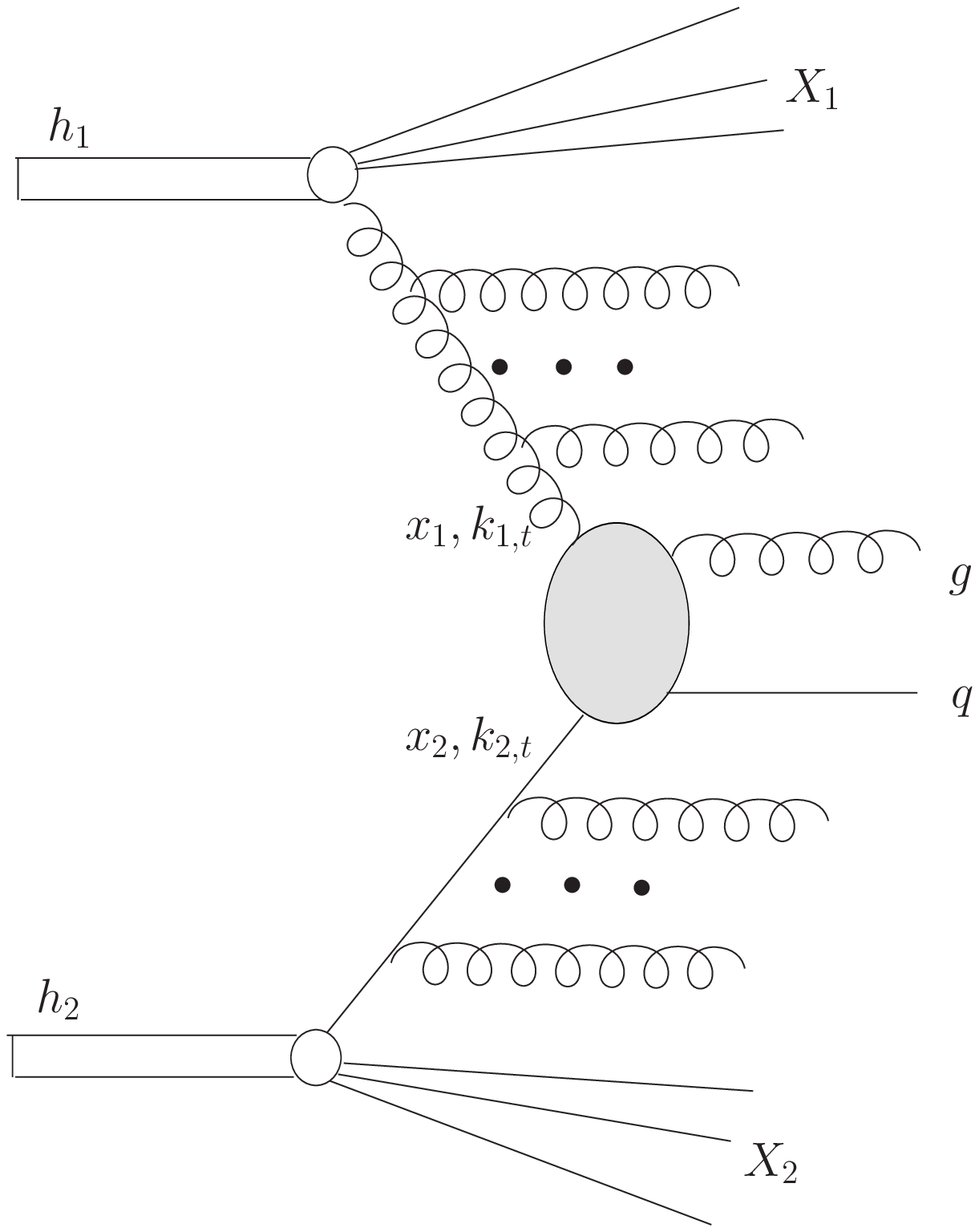} }
\resizebox{0.35\columnwidth}{!}{%
\includegraphics{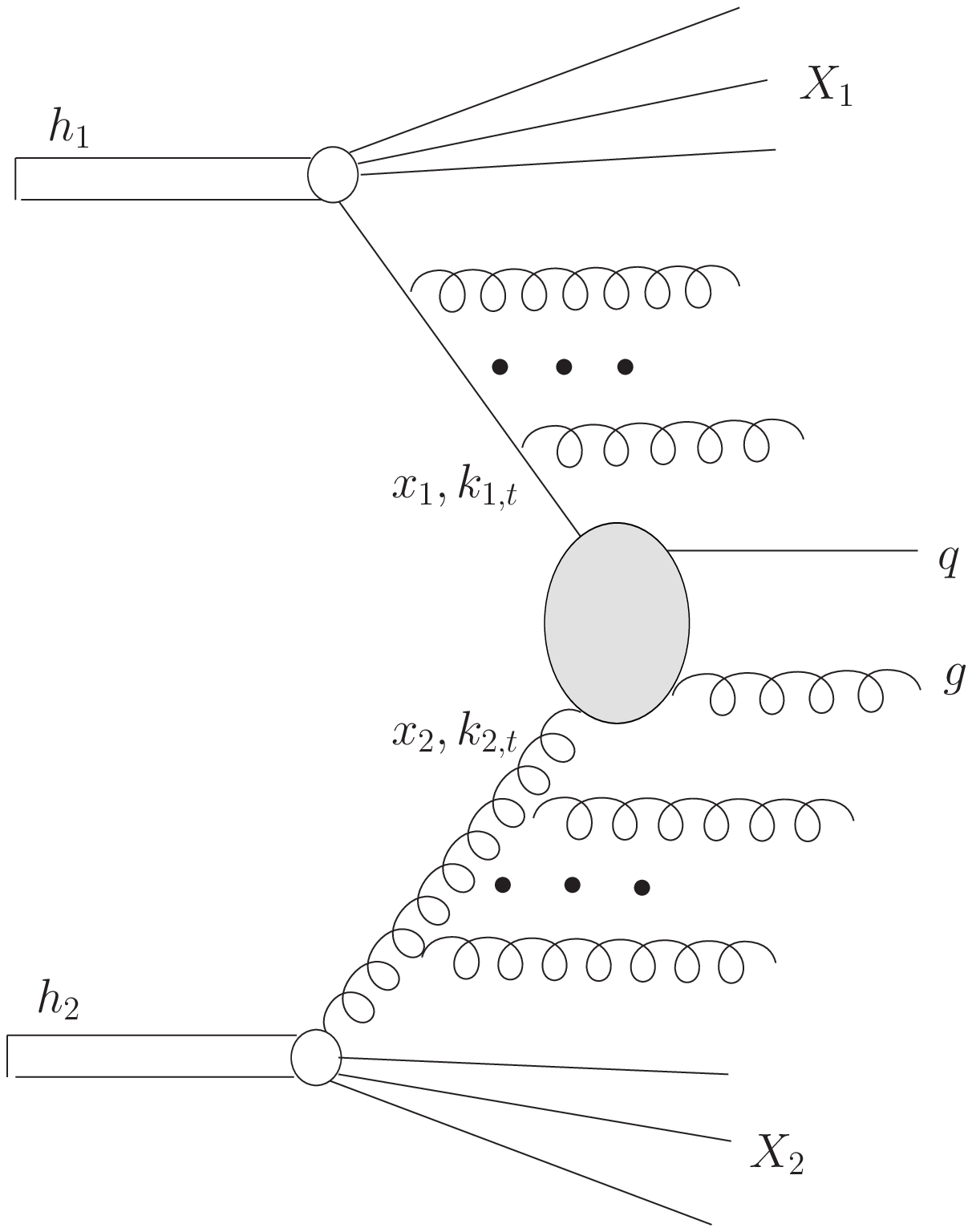} }
\caption{Diagrams for $k_t$-factorization approach to dijet
production.}
\label{fig:kt_factorization_dijets_diagrams}
\end{center}
\end{figure}



\begin{figure}    
\begin{center}
\resizebox{0.35\columnwidth}{!}{%
\includegraphics{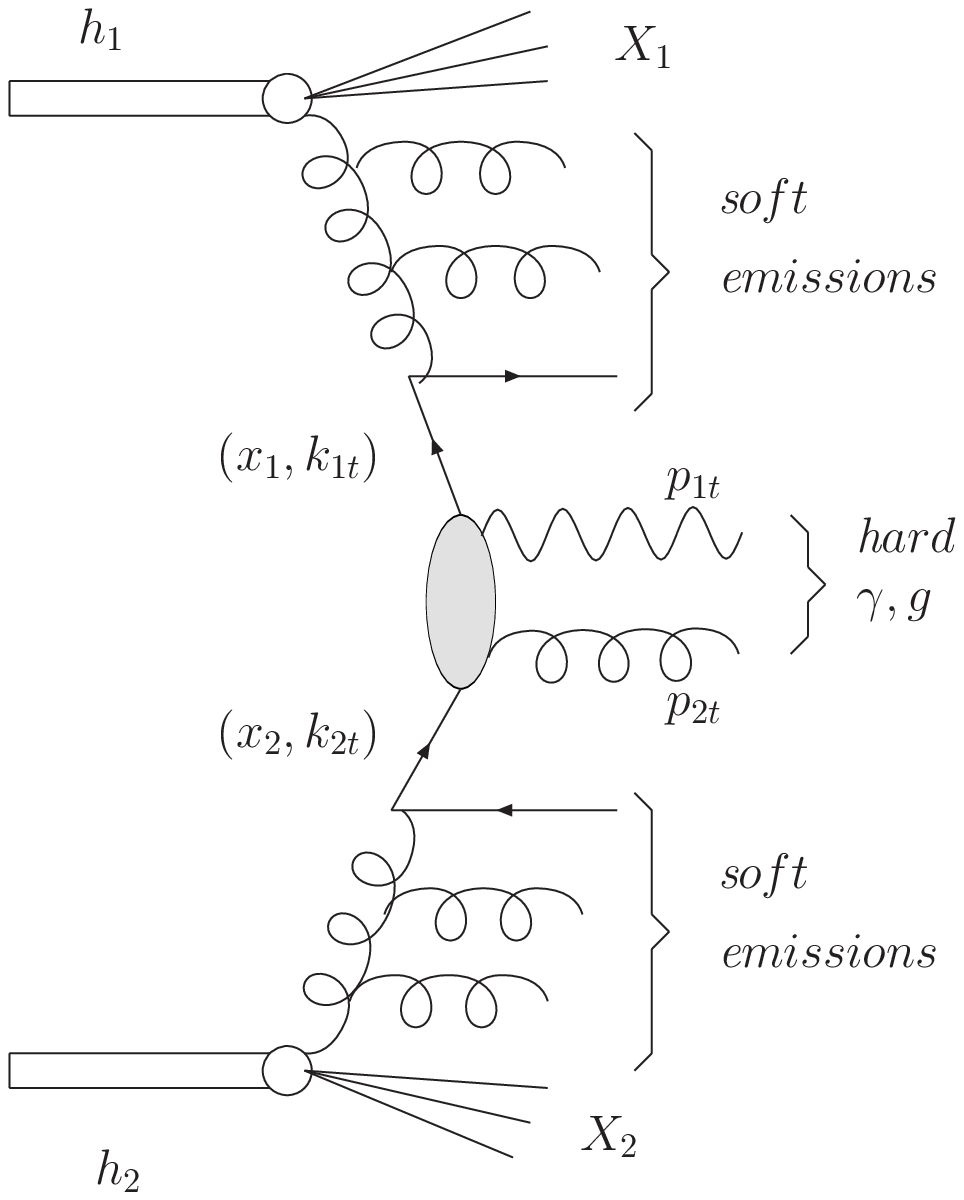} }
\resizebox{0.35\columnwidth}{!}{%
\includegraphics{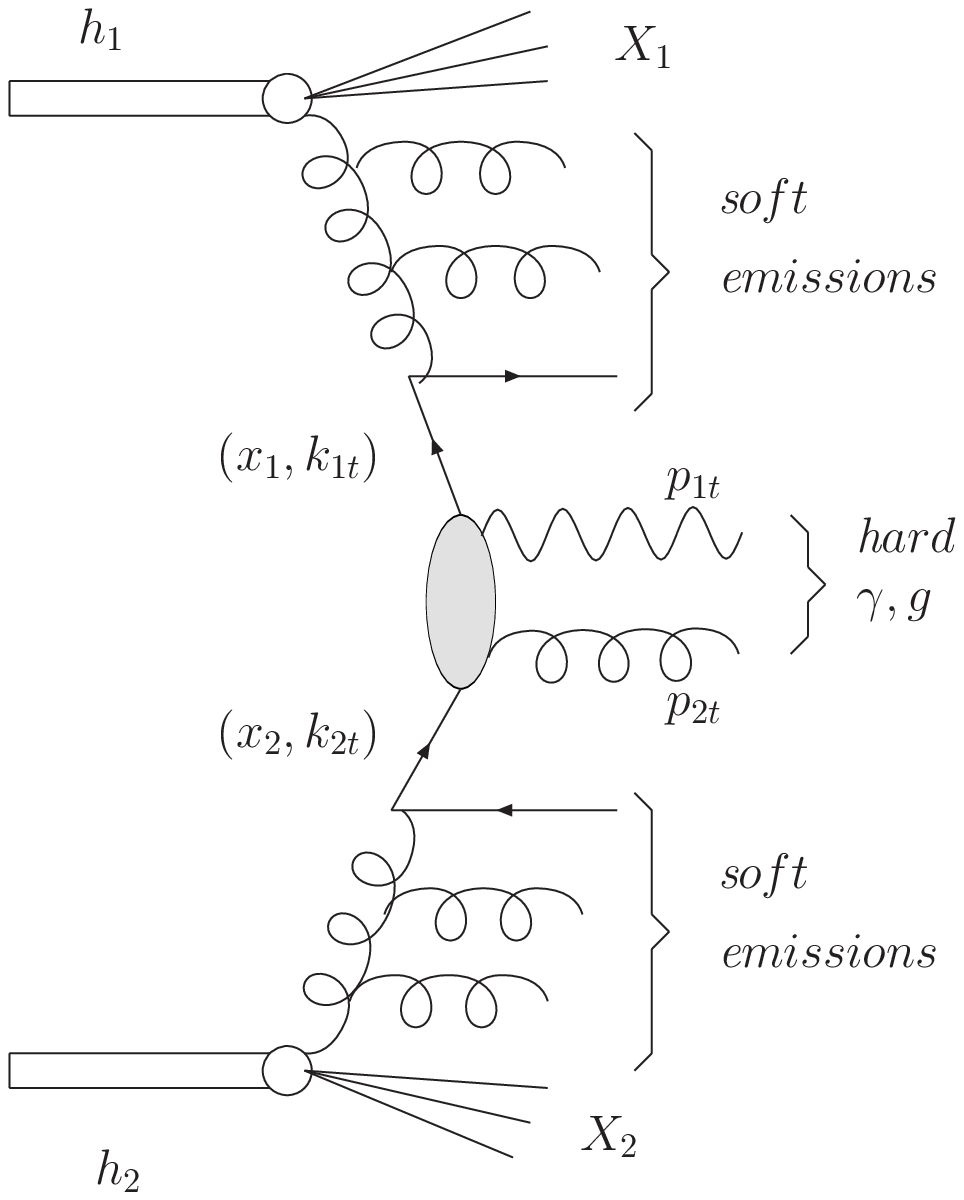} }
\resizebox{0.35\columnwidth}{!}{%
\includegraphics{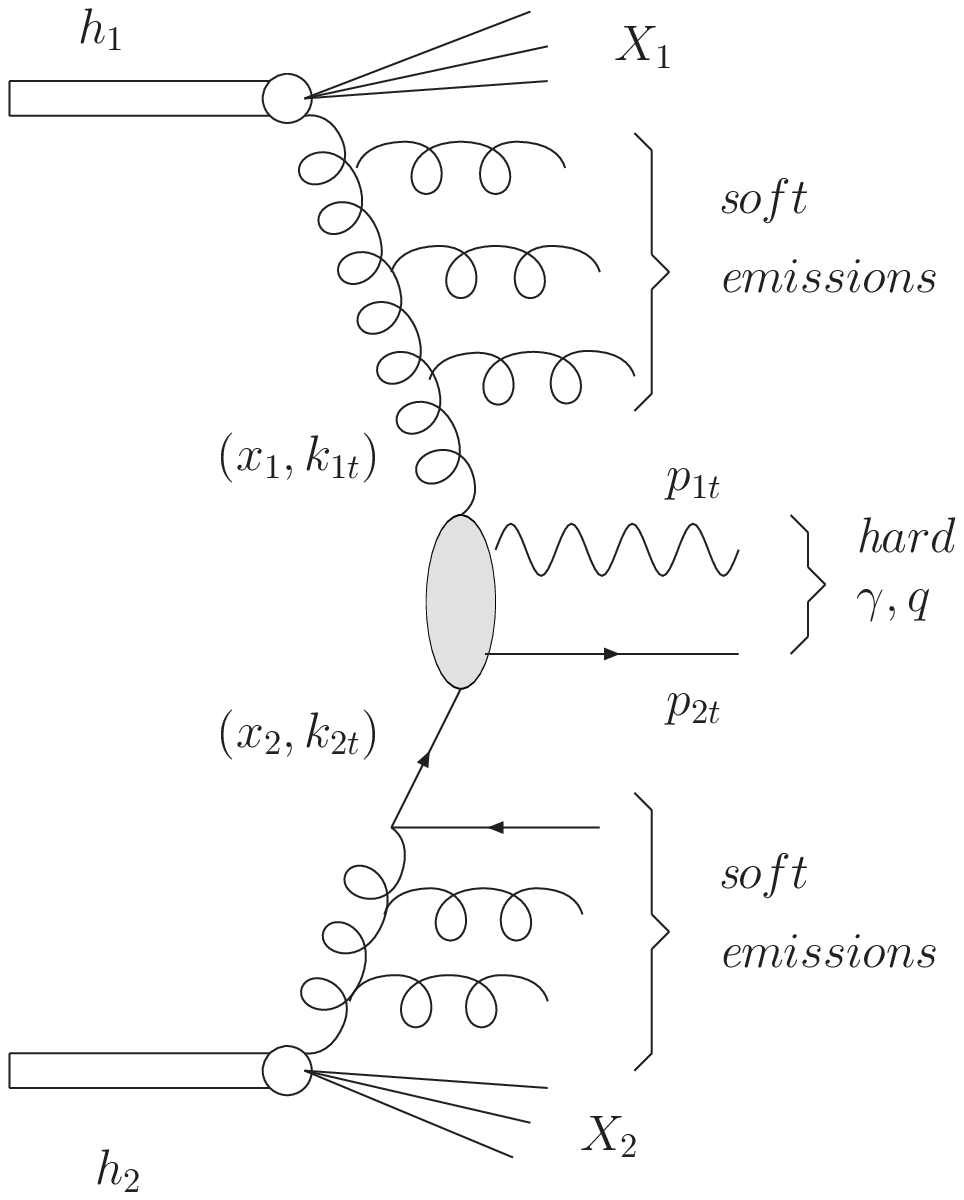} }
\resizebox{0.35\columnwidth}{!}{%
\includegraphics{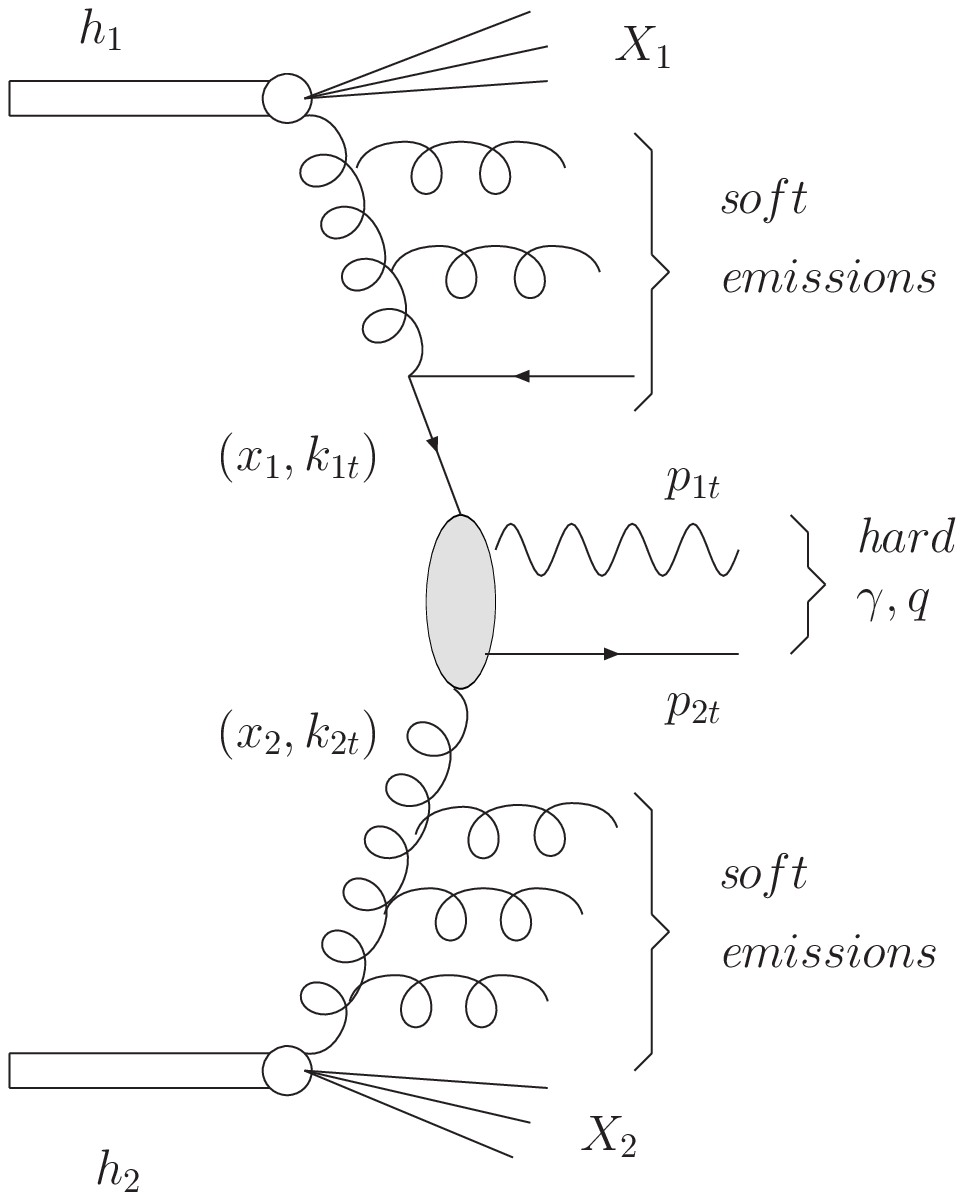} }
\caption{
Diagrams for $k_t$-factorization approach to photon-jet
correlations.}
\label{fig:kt_factorization_photonjet_diagrams}
\end{center}
\end{figure}


It is known that at high energies, at midrapidities and not too large
transverse momenta of the jet (or photon) production is dominated by
(sub)processes initiated by gluons.
In this paper we concentrate on such processes.
In this presentation we discuss mainly $k_t$-factorization approach.
Some aspects of the standard collinear approach are discussed
in Refs.\cite{SRS07,PS07}.

In the $k_t$-factorization approach the cross section for the production
of a pair of partons or photon and parton (k,l) can be written as
\begin{eqnarray}
\frac{d\sigma(h_1 h_2 \rightarrow jet(\gamma) jet)}
{d^2p_{1,t}d^2p_{2,t}} &=& \sum_{i,j,k,l} \int dy_1 dy_2
\frac{d^2 k_{1,t}}{\pi}\frac{d^2 k_{2,t}}{\pi}
\frac{1}{16\pi^2(x_1x_2s)^2}
\overline{|{\cal M}(i j \rightarrow k l)|^2}
\nonumber \\
&\times&\delta^2(\overrightarrow{k}_{1,t}
+\overrightarrow{k}_{2,t}
-\overrightarrow{p}_{1,t}
-\overrightarrow{p}_{2,t})
{\cal F}_i(x_1,k_{1,t}^2){\cal F}_j(x_2,k_{2,t}^2) \; ,
\label{basic_formula}
\end{eqnarray}
where
\begin{equation}
x_1 = \frac{m_{1,t}}{\sqrt{s}}\mathrm{e}^{+y_1} 
    + \frac{m_{2,t}}{\sqrt{s}}\mathrm{e}^{+y_2} \; ,
\end{equation}
\begin{equation}
x_2 = \frac{m_{1,t}}{\sqrt{s}}\mathrm{e}^{-y_1} 
    + \frac{m_{2,t}}{\sqrt{s}}\mathrm{e}^{-y_2} \; ,
\end{equation}
and $m_{1,t}$ and $m_{2,t}$ are so-called transverse masses
defined as $m_{i,t} = \sqrt{p_{i,t}^2+m^2}$, where $m$ is the mass
of a parton.
In the case of photon-jet correlations there is no sum over $k$ 
($k = \gamma$).
In the following we shall assume that all partons are massless.
The objects denoted by ${\cal F}_i(x_1,k_{1,t}^2)$ and
${\cal F}_j(x_2,k_{2,t}^2)$ in the equation above are the unintegrated
parton distributions in hadron $h_1$ and $h_2$, respectively.
They are functions of longitudinal momentum fraction and transverse
momentum of the incoming (virtual) parton.


In Fig.\ref{fig:kt_factorization_dijets_diagrams} we show the diagrams
included for dijet correlations in Ref. \cite{SRS07}.

In Fig.\ref{fig:kt_factorization_photonjet_diagrams} we show similar
diagrams included for photon-jet correlations in Ref. \cite{PS07}.

The formula (\ref{basic_formula}) allows to study different types
of correlations. Here we shall limit to a few examples.
The details concerning unintegrated gluon (parton) distributions
can be found in original publications (see \cite{SRS07,PS07} and
references therein).

\section{Results}
\label{Results}


\begin{figure}    
\begin{center}
\resizebox{0.45\columnwidth}{!}{%
\includegraphics{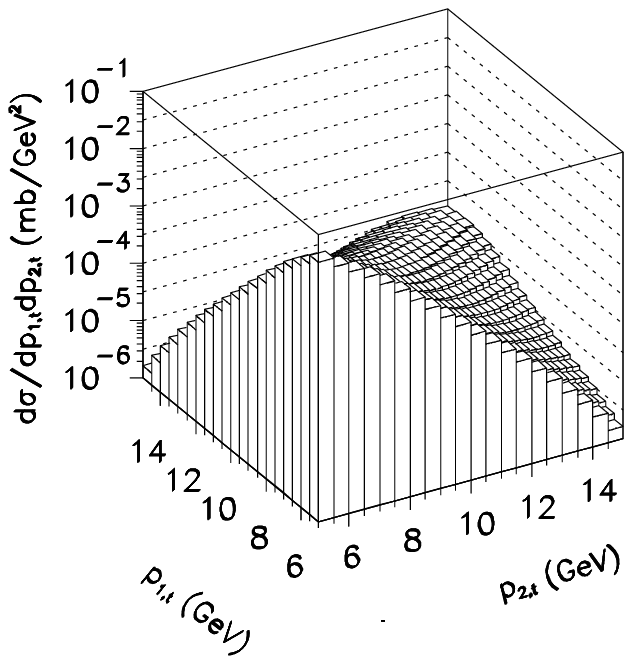}}
\resizebox{0.45\columnwidth}{!}{%
\includegraphics{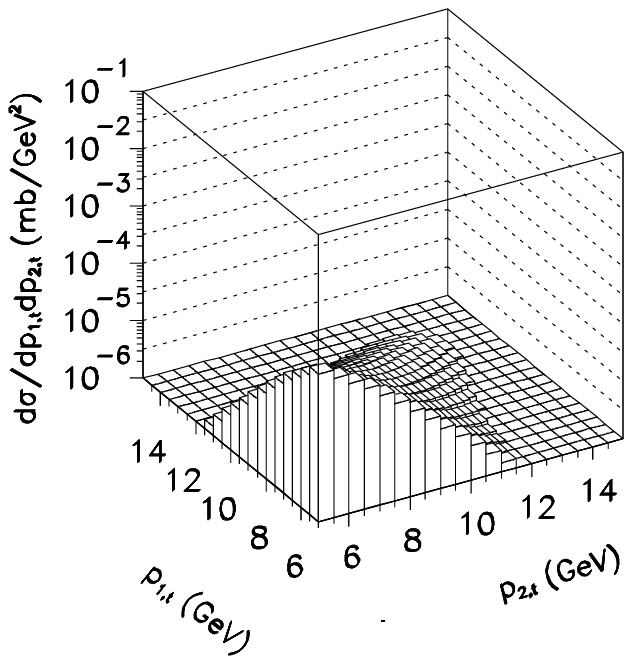}}
\resizebox{0.45\columnwidth}{!}{%
\includegraphics{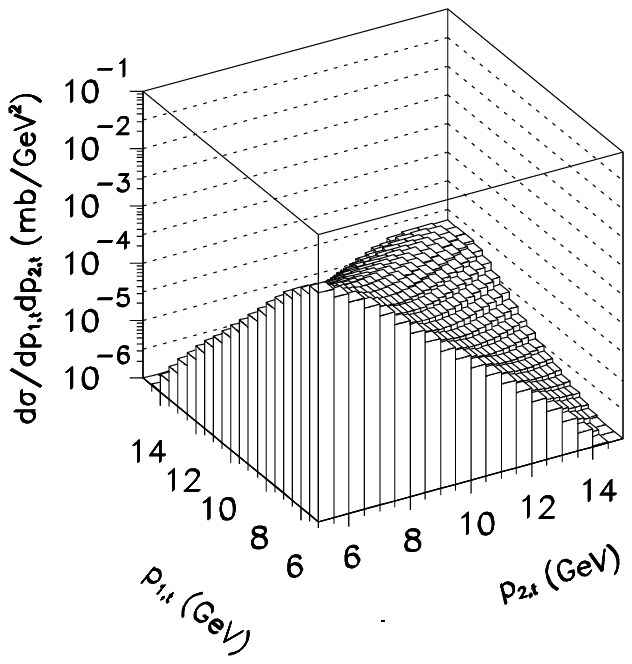}}
\resizebox{0.45\columnwidth}{!}{%
\includegraphics{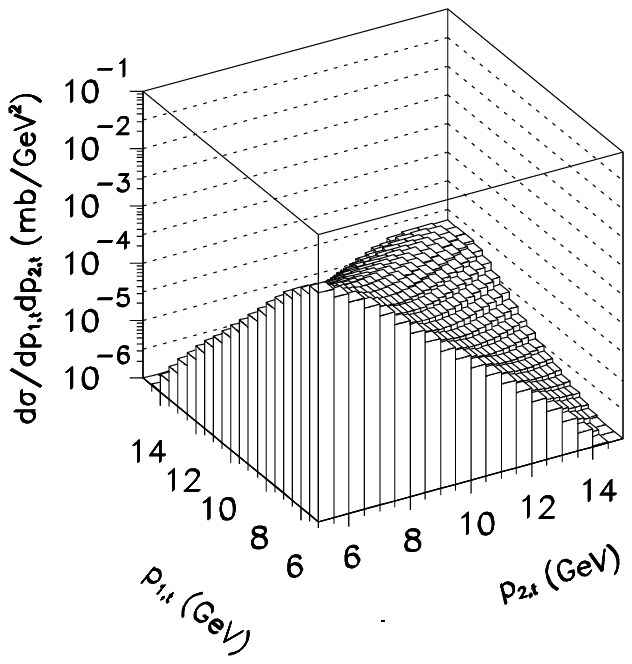}}
\caption{
Two-dimensional distributions in $p_{1,t}$ and $p_{2,t}$ 
for different subprocesses $gg \to gg$ (upper left)
$gg \to q \bar q$ (upper right), $gq \to gq$ (lower left)
and $qg \to qg$ (lower right). In this calculation $\sqrt{s}$ = 200 GeV 
and Kwieci\'nski UPDFs with exponential nonperturbative form factor
($b_0$ = 1 GeV$^{-1}$) and $\mu^2$ = 100 GeV$^2$ were used.
Here integration over full range of parton rapidities was made.}
\label{fig:p1tp2t_dijets_abcd}
\end{center}
\end{figure}


\subsection{Dijet correlations}
\label{Results:dijets}

In Fig.\ref{fig:p1tp2t_dijets_abcd} we show two-dimensional
maps of the cross section in $(p_{1,t},p_{2,t})$ for processes
shown in Fig.\ref{fig:kt_factorization_dijets_diagrams}.
Only very few approaches in the literature include both gluons and
quarks and antiquarks.
In the calculation above we have used Kwieci\'nski UPDFs with
exponential nonperturbative form factor \footnote{ $F(b) = \exp(-b^2/4
  b_0^2)$ multiplies UPDFs in the impact parameter space and
is responsible for nonperturbative effects included in addition to
perturbative effects embedded in the Kwieci\'nski evolution equations (for
more details see e.g. \cite{CS05}).}
($b_0$ = 1 GeV$^{-1}$),
and the factorization scale $\mu^2 = (p_{t,min}+p_{t,max})^2/4$ = 100 GeV$^2$.

\begin{figure}   
\begin{center}
\resizebox{0.65\columnwidth}{!}{%
\includegraphics{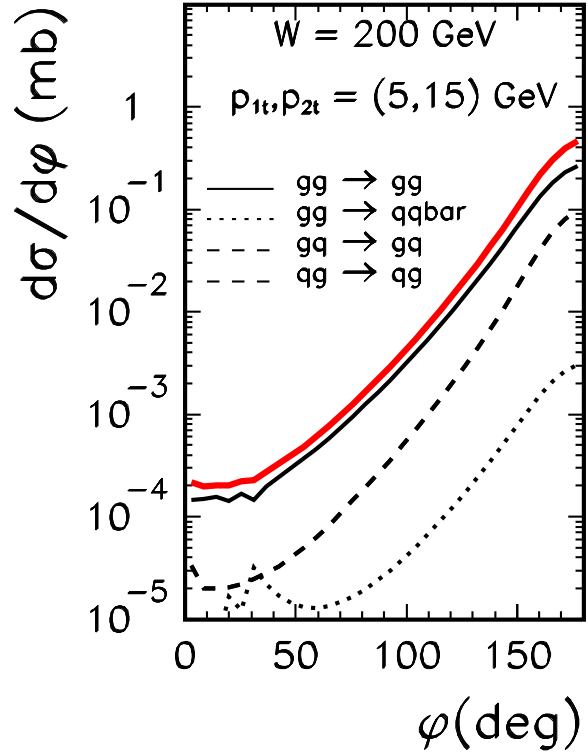}}
\caption{  
The angular correlations for all four components: $gg \to gg$ (solid),
$gg \to q \bar q$ (dashed) and $gq \to gq$ = $qg \to qg$ (dash-dotted).
The calculation is performed with the Kwieci\'nski UPDFs and
$b_0$ = 1 GeV$^{-1}$.
The integration is made for jets from the transverse momentum interval:
5 GeV $< p_{1,t}, p_{2,t} <$ 15 GeV and from the rapidity interval:
-4 $< y_1, y_2 <$ 4.
}
\label{fig:2to2_contributions_phi}
\end{center}
\end{figure}

For completeness in Fig.\ref{fig:2to2_contributions_phi} we show
azimuthal angle dependence of the cross section for all four components.
There is no sizable difference in the shape of azimuthal distribution
for different components.


The Kwieci\'nski approach allows to separate the unknown perturbative
effects incorporated via nonperturbative form factors
and the genuine effects of QCD evolution.
The Kwieci\'nski distributions have two external parameters:
\begin{itemize}
\item the parameter $b_0$ responsible for nonperturbative effects,
such as primordial distribution of partons in the nucleon,
\item the evolution scale $\mu^2$ responsible for the soft
resummation effects.
\end{itemize}

\begin{figure}
\begin{center}
\resizebox{0.65\columnwidth}{!}{%
\includegraphics{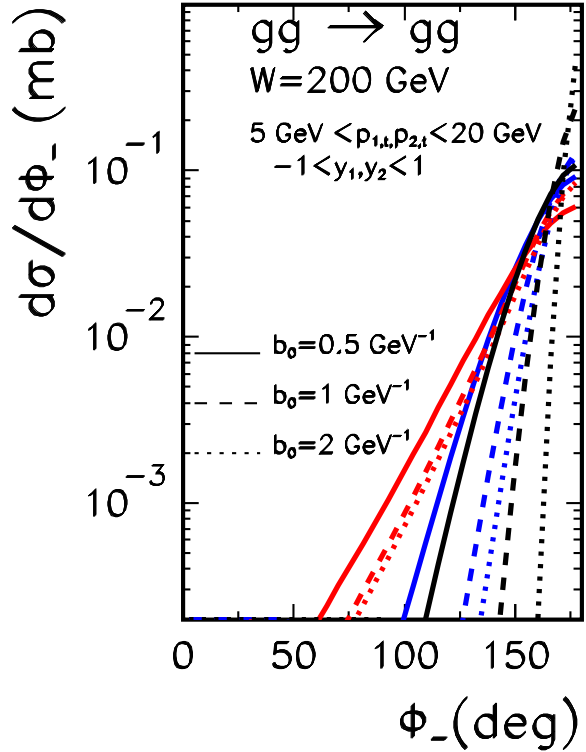}}
\caption{
The azimuthal correlations for the $gg \to gg$ component obtained with
the Kwieci\'nski UGDFs for different values of the nonperturbative
parameter $b_0$ and for different evolution scales $\mu^2$ = 10 
(grey, blue online), 100 (black, red online) GeV$^2$.
The initial distributions (without evolution) are shown for reference
by black lines.
}
\label{fig:b0_mu2_phi}
\end{center}
\end{figure}

While the latter can be identified physically with characteristic
kinematical quantities in the process $\mu^2 \sim p_{1,t}^2, p_{2,t}^2$,
the first one is of nonperturbative origin and cannot be calculated
from first principles.
The shapes of distributions depends, however, strongly on the value of
the parameter $b_0$.
This is demonstrated in Fig.\ref{fig:b0_mu2_phi} for the $gg \to gg$
subprocess.
The smaller $b_0$ the bigger decorrelation in azimuthal angle
can be observed. In Fig.\ref{fig:b0_mu2_phi} we show also the role of
the evolution scale in the Kwieci\'nski distributions.
The QCD evolution embedded in the Kwieci\'nski evolution equations
populate larger transverse momenta of partons entering the hard process.
This significantly increases the initial (nonperturbative) decorrelation
in azimuth.
For transverse momenta of the order of $\sim$ 10 GeV the effect of
evolution is of the same order of magnitude as the effect characteristic
for the nonperturbative physics. For larger scales of the order of $\mu^2
\sim$ 100 GeV$^2$, more adequate for jet production, the initial
condition is of minor importance and the effect of decorrelation
is dominated by the evolution. Asymptotically (infinite scales) there is
no dependence on the initial condition provided reasonable initial
conditions are taken.


\begin{figure}    
\begin{center}
\resizebox{0.65\columnwidth}{!}{%
\includegraphics{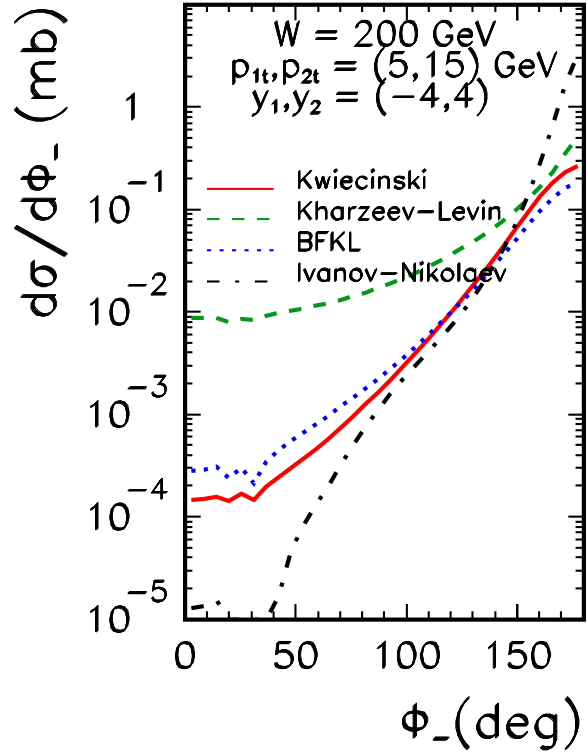}}
\caption{
Azimuthal angle correlations between two jets for different
unintegrated gluon distributions.}
\label{fig:dijets_dsig_dphi_updf}
\end{center}
\end{figure}

In Fig.\ref{fig:dijets_dsig_dphi_updf} we show azimuthal-angle correlations
for the $g g \to g g$ component (dominant at midrapidities)
for different UGDFs from the literature. Rather different results are
obtained for different UGDFs. In principle, experimental results
could select the ``best'' UGDF. We do not need to mention that such
measurements are not easy at RHIC and hadron correlations are studied
instead of jet correlations.


\subsection{Photon-jet correlations}
\label{Results:photon-jet}

\begin{figure} 
\begin{center}
\resizebox{0.45\columnwidth}{!}{%
\includegraphics{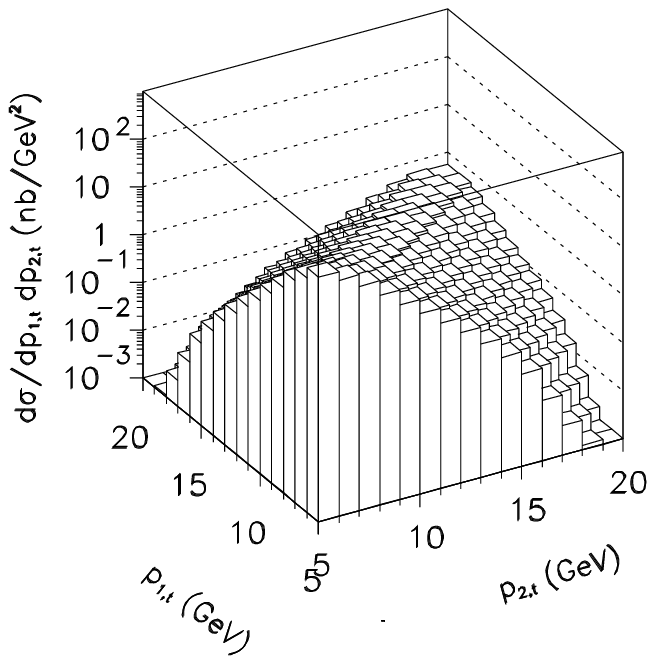}}
\resizebox{0.45\columnwidth}{!}{%
\includegraphics{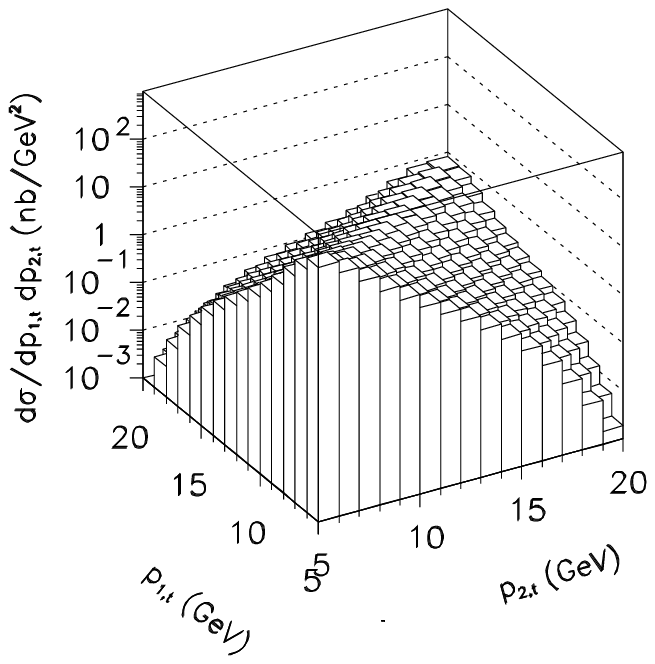}}
\resizebox{0.45\columnwidth}{!}{%
\includegraphics{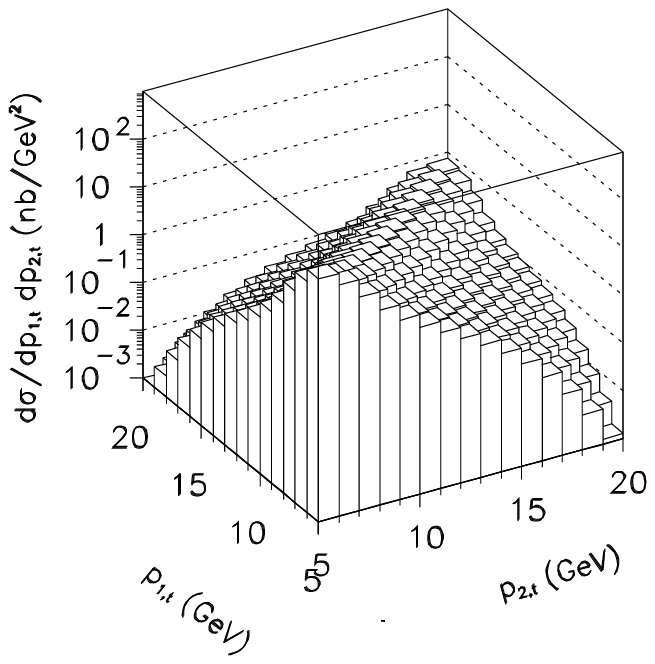}}
\resizebox{0.45\columnwidth}{!}{%
\includegraphics{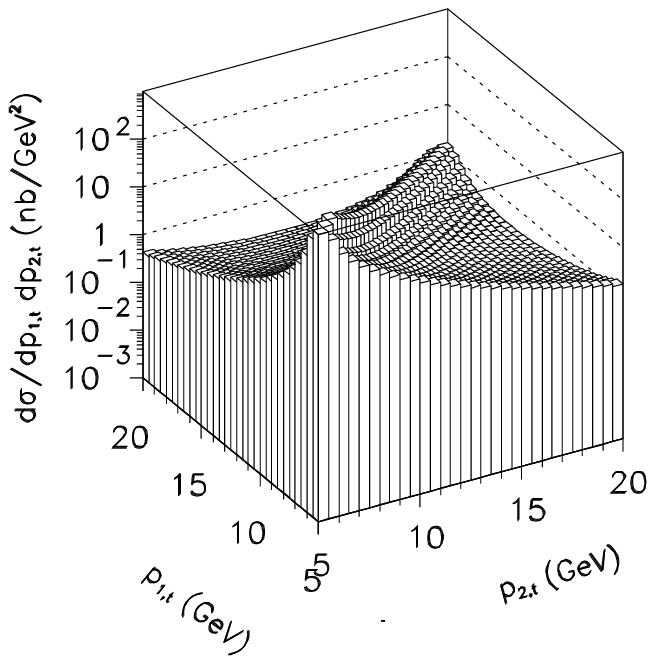}}
\caption{
Transverse momentum distributions $d\sigma/dp_{1,t}dp_{2,t}$
at $\sqrt{s}$ = 1960 GeV and  
for different UPDFs in the $k_t$-factorization approach for
Kwieci\'nski ($b_0$ = 1 GeV$^{-1}$, $\mu^2$ = 100 GeV$^{2}$) (upper left),
BFKL (upper right), KL (lower left) 
and NLO $2 \to 3$ collinear-factorization approach (lower right).
The integration over rapidities from the interval -5 $< y_1, y_2 <$
5 is performed.
}
\label{fig:updfs_ptpt}
\end{center}
\end{figure}

Let us start from presenting our results on the $(p_{1,t},p_{2,t})$ plane.
In Fig.\ref{fig:updfs_ptpt} we show the maps for different 
UPDFs used in the $k_t$-factorization approach as well as for NLO 
collinear-factorization approach for
$p_{1,t}, p_{2,t} \in (5,20)$~GeV and at the Tevatron energy $\sqrt(s) =
1960$~GeV. In the case of the Kwieci\'nski distribution we have taken
$b_0$ = 1 GeV$^{-1}$ for the exponential nonperturbative form factor
and the scale parameter $\mu^2$ = 100 GeV$^2$.
Rather similar distributions are obtained for different UPDFs.
The distribution obtained in the NLO approach differs qualitatively
from those obtained in the $k_t$-factorization approach.
First of all, one can see a sharp ridge along the diagonal $p_{1,t} = p_{2,t}$.
This ridge corresponds to a soft singularity when the unobserved
parton has very small transverse momentum $p_{3,t}$.
At the same time this corresponds to the azimuthal
angle between the photon and the jet being $\phi_{-} = \pi$. Obviously this is
a region which cannot be reliably calculated in collinear pQCD.
There are different practical possibilities to exclude this region from
the calculations \cite{PS07}.

\begin{figure} 
\begin{center}
\resizebox{0.65\columnwidth}{!}{%
\includegraphics{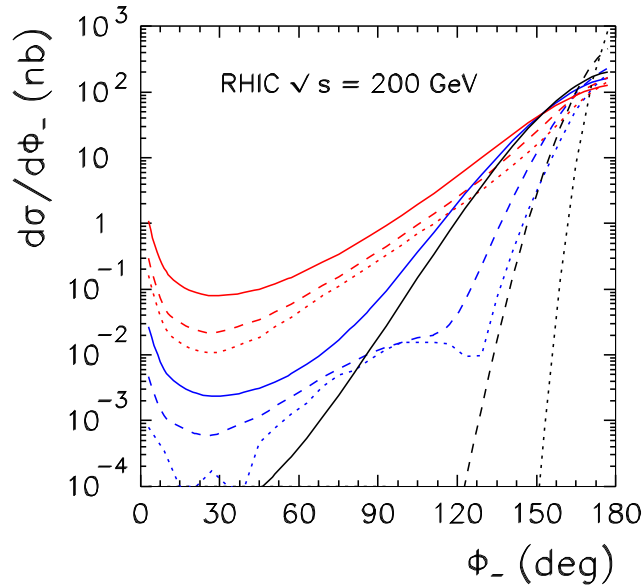}}
\caption{
(Color online) Azimuthal angle correlation functions at RHIC
energies for different scales and different values of $b_0$ 
of the Kwieci\'nski distributions.
The solid line is for $b_0$ = 0.5 GeV$^{-1}$, the dashed line is for
$b_0$ = 1 GeV$^{-1}$ and the dotted line is for $b_0$ = 2 GeV$^{-1}$.
Three different values of the scale parameters are shown: 
$\mu^2$ = 0.25, 10, 100 GeV$^2$ (the bigger the scale the bigger
the decorellation effect, different colors on line).
In this calculation  $p_{1,t}, p_{2,t} \in$ (5,20) GeV and
$y_1, y_2 \in$ (-5,5).
}
\label{fig:photon_jet_kwiecinski_scale}
\end{center}
\end{figure}

As discussed in Ref.\cite{PS06_photon} the Kwieci\'nski distributions
are very useful to treat both the nonperturbative (intrinsic
nonperturbative transverse momenta)
and the perturbative (QCD broadening due to parton emission) effects on
the same footing.
In Fig.\ref{fig:photon_jet_kwiecinski_scale} we show the effect of
the scale evolution of the Kwieci\'nski UPDFs on the azimuthal angle
correlations between the photon and the associated jet.
We show results for different initial conditions ($b_0$ = 0.5, 1.0, 2.0
GeV$^{-1}$). At the initial scale (fixed here as in the original
GRV \cite{GRV98} to be $\mu^2$ = 0.25 GeV$^2$) there is a sizable
difference of the results for different $b_0$. The difference
becomes less and less pronounced when the scale increases.
At $\mu^2$ = 100 GeV$^2$ the differences practically disappear.
This is due to the fact that the QCD-evolution broadening of
the initial parton transverse momentum distribution is much bigger than
the typical initial nonperturbative transverse momentum scale.

\begin{figure} %
\begin{center}
\resizebox{0.65\columnwidth}{!}{%
\includegraphics{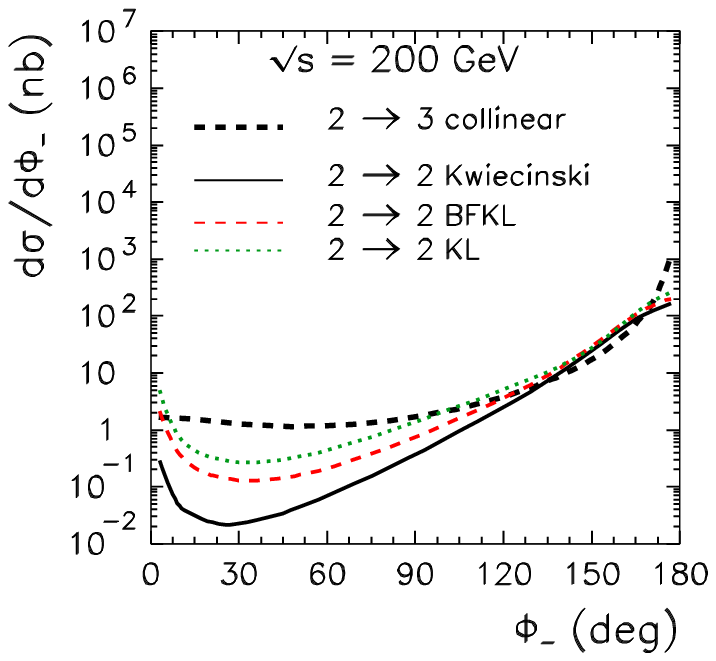}}
\caption{
Photon-jet angular azimuthal correlations $d\sigma/d\phi_-$
for proton-(anti)proton collision at $\sqrt{s}$ = 200~GeV  
for different UPDFs in the $k_t$-factorization approach for
the Kwieci\'nski (solid), BFKL (dashed), KL (dotted) UPDFs/UGDFs
and for the NLO collinear-factorization approach (thick dashed).
Here $y_1, y_2 \in (-5,5)$.
}
\label{fig:updfs_phid}
\end{center}
\end{figure}

In Fig.\ref{fig:updfs_phid} we show azimuthal angular
correlations for RHIC.
In this case integration is made over transverse momenta 
$p_{1,t}, p_{2,t} \in (5,20)$~GeV and rapidities $y_1, y_2 \in
(-5,5)$. The standard NLO collinear cross section grows somewhat faster
with energy than the $k_t$-result with unintegrated Kwieci\'nski
distribution. This is partially due to approximation made in calculation
of the off-shell matrix elements.

\begin{figure} 
\begin{center}
\resizebox{0.65\columnwidth}{!}{%
\includegraphics{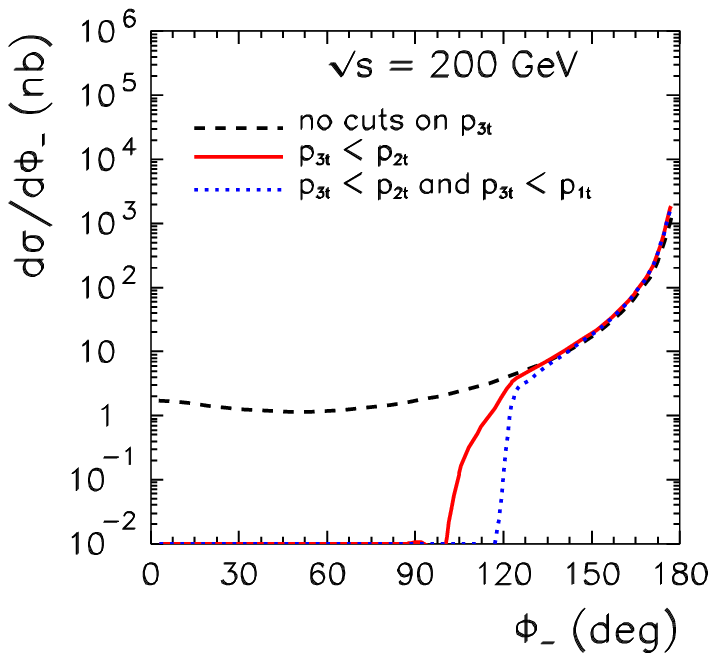}}
\caption{
Angular azimuthal correlations for different cuts on the transverse 
momentum of third (unobserved) parton in the NLO collinear-factorization
approach without any extra constraints (dashed), $p_{3,t} < p_{2,t}$ (solid),
$p_{3,t} < p_{2,t}$ and $p_{3,t} < p_{1,t}$ in addition (dotted).
Here $\sqrt{s} = 200$~GeV and $y_1, y_2 \in (-5,5)$.
}
\label{fig:leading_jets_angle}
\end{center}
\end{figure}

Let us consider now some aspect of the standard NLO approach.
Here 3 jets with transverse momenta $p_{1,t}, p_{2,t}$ and $p_{3,t}$
are produced \footnote{Jet 1 (with $p_{1,t}$) and jet 2 (with $p_{2,t}$)
are those which correlations are studied.}.
In Fig.\ref{fig:leading_jets_angle} we show angular azimuthal
correlations for different interrelations between transverse momenta
of outgoing photon and partons:
(a) with no constraints on $p_{3,t}$, (b) the case where $p_{2,t} >
p_{3,t}$ condition (called leading jet condition in the following)
is imposed, (c) $p_{2,t} > p_{3,t}$ and
an additional condition $p_{1,t} > p_{3,t}$.
The results depend significantly on the scenario chosen as can be seen
from the figure. The general pattern is very much the same for different
energies.
The figure demonstrates that only higher-orders contribute to the region
of small relative angles.
The same is true for dijet correlations discussed in Ref.\cite{SRS07}.
We wish to notice that there are no such limitations in the 
$k_t$-factorization approach which implicitly include the higher orders.

\section{Conclusions}
\label{conclusions}

Motivated by the recent experimental results of hadron-hadron correlations
at RHIC we have discussed jet-jet and photon-jet correlations.

In comparison to recent works on dijet production in the framework of
$k_t$-factorization approach, we have included two new mechanisms based
on $gq \to gq$ and $qg \to qg$ hard subprocesses.
This was done using the Kwieci\'nski unintegrated parton
distributions.
We find that the new terms give significant contribution at RHIC energies.
In general, the results of the $k_t$-factorization approach depend
on UGDFs/UPDFs used, i.e. on approximation and assumptions made
in their derivation.

An interesting observation has been made for azimuthal angle correlations.
At relatively small transverse momenta ($p_t \sim$ 5--10 GeV)
the $2 \to 2$ subprocesses, not contributing to the correlation
function in the collinear approach, dominate over $2 \to 3$ components.
The latter dominate only at larger transverse momenta, i.e. in
the traditional jet region.

The results obtained in the standard NLO approach depend significantly
whether we consider correlations of any jets or correlations of only
leading jets. 
In the NLO approach one obtains
$\frac{d \sigma}{d \phi_{-}}$ = 0 if $\phi_{-} < \frac{2}{3} \pi$
for leading jets as a result of a kinematical constraint.
Similarly $\frac{d \sigma}{d p_{1,t} d p_{2,t}}$ = 0 if $p_{1,t} > 2
p_{2,t}$ or $p_{2,t} > 2 p_{1,t}$. In this presentation we have
discussed explicitly only a similar case of photon-jet correlations.

There is no such a constraint in the $k_t$-factorization approach
which gives a nonvanishing cross section at small relative azimuthal
angles between leading jets and transverse-momentum asymmetric
configurations. We conclude that in these regions the $k_t$-factorization
approach is a good and efficient tool for the description of leading-jet
correlations.
Rather different results are obtained with different UGDFs
which opens a possibility to verify them experimentally.

On the contrary, in the case of correlations of any unrestricted jets
(all possible dijet combinations)
the NLO cross section exceeds the cross section obtained in
the $k_t$-factorization approach with different UGDFs. This is therefore
a domain of the standard fixed-order pQCD.
We recommend such an analysis as an alternative to study leading-jet
correlations.

Consequences for particle-particle correlations, measured
recently at RHIC, require a separate dedicated analysis.

We have discussed also photon-jet correlation observables.
Up to now such correlations have not been studied experimentally.
As for the dijet case we have concentrated on the
region of small transverse momenta (semi-hard region) where
the $k_t$-factorization approach seems to be the most efficient and
theoretically justified tool.
We have calculated correlation observables for different unintegrated parton
distributions from the literature. Our previous analysis of
inclusive spectra of direct photons suggests that the Kwieci\'nski
distributions give the best description at low and intermediate
energies.
We have discussed the role of the evolution scale of the Kwieci\'nski
UPDFs on the azimuthal correlations. In general, the bigger the scale
the bigger decorrelation in azimuth is observed. When the scale
$\mu^2 \sim p_t^2$(photon) $\sim p_t^2$(associated jet)
(for the  kinematics chosen $\mu^2 \sim$ 100 GeV$^2$) is assumed, much bigger 
decorrelations can be observed than from the standard Gaussian smearing
prescription often used in phenomenological studies.

The correlation function depends strongly on whether it is the
correlation of the
photon and any jet or the correlation of the photon and the leading-jet
which is considered.
In the last case there are regions in azimuth and/or in the two-dimensional
($p_{1,t}, p_{2,t}$) space which cannot be populated in the standard
next-to-leading order approach. In the latter case the $k_t$-factorization
seems to be a useful and efficient tool.

At RHIC one can measure jet-hadron correlations only for not too high
transverse momenta of the trigger photon and of the associated hadron.
This is precisely the semihard region discussed here.
In this case the theoretical calculations would require inclusion of the
fragmentation process. This can be done easily assuming independent parton
fragmentation method.

\vskip 0.5cm

{\bf Acknowledgments.}
This presentation is devoted to the memory of Jozs\'o Zim\'anyi,
fantastic physicist, extraordinary man and very good friend of the
Polish Nuclear and Particle Physics Community.
A.S. congratulates the Budapest group for organizing the fantastic
memorial workshop.
We are also indebted to Jan Rak from the PHENIX collaboration for the
discussion of recent correlation results from RHIC.
This work was partially supported by the grant
of the Polish Ministry of Scientific Research and Information Technology
number 1 P03B 028 28.

\end{document}